# *First Principles Calculation of the Topological Phases of the Photonic Haldane Model*

**Filipa R. Prudêncio[1,2*], and Mário G. Silveirinha[1]**

[1]University of Lisbon – Instituto Superior Técnico and Instituto de Telecomunicações, Avenida Rovisco Pais 1, 1049-001 Lisbon, Portugal

[2]Instituto Universitário de Lisboa (ISCTE-IUL), Avenida das Forças Armadas 376, 1600-077 Lisbon, Portugal

## Abstract

Photonic topological materials with a broken time-reversal symmetry are characterized by nontrivial topological phases, such that they do not support propagation in the bulk region but forcibly support a nontrivial net number of unidirectional edge-states when enclosed by an opaque-type boundary, e.g., an electric wall. The Haldane model played a central role in the development of topological methods in condensed-matter systems, as it unveiled that a broken time-reversal is the essential ingredient to have a quantized electronic Hall phase. Recently, it was proved that the magnetic field of the Haldane model can be imitated in photonics with a spatially varying pseudo-Tellegen coupling. Here, we use a Green's function method to determine from "first principles" the band diagram and the topological invariants of the photonic Haldane model, implemented as a Tellegen photonic crystal. Furthermore, the topological phase diagram of the system is found, and it is shown with first principles calculations that the granular structure of the photonic crystal can create nontrivial phase transitions controlled by the amplitude of the pseudo-Tellegen parameter.

---

1. [*] Corresponding author: filipa.prudencio@lx.it.pt

# I. Introduction

The study of topological properties of physical systems and of how the topology influences the physical responses and phenomena has been a very active field of research in recent years [1-20]. The topology of physical system is typically determined by the global properties of the operator that describes the time evolution of the system state (in quantum systems, this operator is the Hamiltonian). Here, we focus on a particular class of topological systems known as Chern insulators [8]. Such systems have a broken time reversal symmetry and are characterized by a topological invariant known as the Chern number. The key fingerprint of a nontrivial topological phase is the emergence of gapless scattering immune unidirectional edge-states at the boundary of the material. This property makes the response of topological systems rather insensitive to fabrication imperfections, disorder and other perturbations [10].

In the 1980s, Haldane discovered that electronic systems with a broken time-reversal symmetry (e.g., an electron gas biased with a magnetic field) may have a quantized Hall conductivity, even if the spatial-average of the magnetic field vanishes [2]. The Haldane model is essentially a tight-binding description of the propagation of an electron wave in a honeycomb lattice of scattering centers (electric potential interaction), subject also to the influence of a periodic magnetic potential. The periodicity of the magnetic potential ensures that the net magnetic flux in a unit cell vanishes. Sometime ago, it was shown that the model theoretically envisioned by Haldane may be implemented in practice using an "artificial graphene" superlattice biased with a spatially varying magnetic field [21-22]. More recently, Ref. [23] developed a photonic analogue for the Haldane model, with the magnetic field of the original electronic model imitated by a spatially varying pseudo-Tellegen coupling. In Ref. [23], the topological phases of the photonic system were determined relying on a tight-binding approximation. Here, building on these previous works, we use an exact Green's function method [20, 24-26] to calculate from "first principles" the topological invariants of the Haldane photonic crystal.

The article is organized as follows. In Section II we present a brief overview of the electronic Haldane model and of its electromagnetic analogue. In Section III the Green's function formalism is applied to the electronic (tight-binding) Haldane model and to the Haldane photonic crystal formed by materials with a pseudo-Tellegen response. In Section IV the topological phases of the electronic and photonic models are calculated based on a Green's function approach. A short summary of the key results is given in Section V.

## II. The Haldane model

In this Section we briefly review the original Haldane model and its electromagnetic analogue [2, 22, 23].

### A. The electronic Haldane model

The Haldane model [2] describes the electronic stationary states of a hexagonal two-dimensional array of scattering centers under the influence of a static magnetic field with a vanishing net flux [Fig. 1a]. There are two inequivalent scattering centers per unit cell, analogous to graphene. The direct lattice primitive vectors are $\mathbf{a}_1 = a/2(3,-\sqrt{3})$ and $\mathbf{a}_2 = a/2(3,\sqrt{3})$, where $a$ is the distance between nearest neighbors [Fig. 1a]. In a tight-binding approximation, the Hamiltonian of the Haldane model in the spectral domain is determined by the 2 ×2 matrix:

$$\mathbf{H}(\mathbf{k}) = \varepsilon(\mathbf{k})\mathbf{1} + d_x(\mathbf{k})\boldsymbol{\sigma}_x + d_y(\mathbf{k})\boldsymbol{\sigma}_y + d_z(\mathbf{k})\boldsymbol{\sigma}_z. \qquad (1)$$

In the above, $\boldsymbol{\sigma}_i$ are the Pauli matrices and the relevant coefficients are:

$$\varepsilon(\mathbf{k}) = 2t_2 \cos(\phi)\left[\cos(\mathbf{k}\cdot\mathbf{a}_1) + \cos(\mathbf{k}\cdot\mathbf{a}_2) + \cos(\mathbf{k}\cdot(\mathbf{a}_2-\mathbf{a}_1))\right],$$

$$d_x(\mathbf{k}) = t_1\left[2\cos\left(k_x\frac{a}{2}\right)\cos\left(k_y a\frac{\sqrt{3}}{2}\right) + \cos(k_x a)\right],$$

$$d_y(\mathbf{k}) = t_1\left[2\sin\left(k_x\frac{a}{2}\right)\cos\left(k_y a\frac{\sqrt{3}}{2}\right) - \sin(k_x a)\right],$$

$$d_z(\mathbf{k}) = M - 2t_2 \sin(\phi)\left[\sin(\mathbf{k}\cdot\mathbf{a}_1) - \sin(\mathbf{k}\cdot\mathbf{a}_2) + \sin(\mathbf{k}\cdot(\mathbf{a}_2-\mathbf{a}_1))\right].$$

(2)

Here, $t_1$ and $t_2$ are the nearest-neighbors and next-nearest neighbors hopping energies, respectively, $M$ is the so-called "mass term" and $\phi$ is the phase factor determined by the coupling between next-nearest neighbors due to the applied magnetic potential. The parameter $M$ is non-zero when the scattering centers associated with the different sub-lattices are different.

The Hamiltonian $\mathbf{H}(\mathbf{k})$ acts on a two-component pseudo-spinor. The components of the pseudo-spinor represent the (Fourier) coordinates of the wave function in the tight-binding basis. The energy dispersion is determined by $\det(\mathbf{H}-\mathcal{E}\mathbf{1})=0$ which yields exactly two electronic bands: $\mathcal{E}_\mathbf{k}^\pm(\mathbf{k}) = \varepsilon \pm \sqrt{d_x^2 + d_y^2 + d_z^2}$. The wave vector is defined over the first Brillouin zone. The primitive vectors of the reciprocal lattice are:

$$\mathbf{b}_1 = \frac{2\pi}{a}\left(\frac{1}{3}, -\frac{1}{\sqrt{3}}\right), \quad \mathbf{b}_2 = \frac{2\pi}{a}\left(\frac{1}{3}, \frac{1}{\sqrt{3}}\right). \tag{3}$$

The energy bands $\mathcal{E}_\mathbf{k}^\pm(\mathbf{k})$ can touch only if $d_x(\mathbf{k}) = d_y(\mathbf{k}) = d_z(\mathbf{k}) = 0$. The only points of the Brillouin zone that may satisfy $d_x(\mathbf{k}) = d_y(\mathbf{k}) = 0$ are the two high symmetry (Dirac) points:

$$K = \frac{2\pi}{3a}\left(1, \frac{1}{\sqrt{3}}\right), \quad K' = \frac{2\pi}{3a}\left(1, -\frac{1}{\sqrt{3}}\right). \tag{4}$$

Thus, in the Haldane model the two bands are typically separated by a complete band gap unless they touch at one of the Dirac points. Below, we show that for large values of $t_2$ the bandgap may be closed, even if the two bands do not intersect.

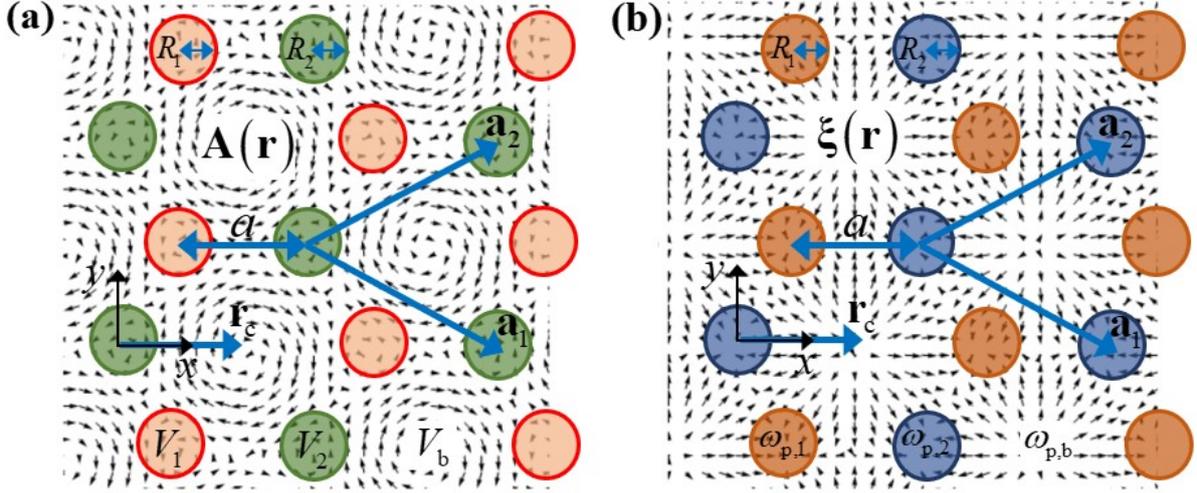

**Fig. 1** Illustrative geometry of **(a)** the electronic Haldane model and **(b)** the equivalent photonic structure. **a)** Representation of the lines of the periodic vector potential $\mathbf{A}(\mathbf{r})$ (black arrows). The disks indicate the location of the scattering centers associated with the electrostatic potentials $V_i$ (i=1,2). In the background region the potential is $V_b$. The distance between nearest neighbors is $a$. **(b)** Representation of the lines of the pseudo-Tellegen vector $\xi(\mathbf{r})$ (black arrows) that determines the nonreciprocal coupling. The plasma frequencies in the scattering centers and in the background are $\omega_{p,i}$ (i=1,2) and $\omega_{p,b}$, respectively.

Reference [22] introduced a possible physical realization of the Haldane model. It relies on a 2D electron gas superlattice patterned with scattering centers, whose effect is modelled by a periodic electric potential $V(\mathbf{r})$. The electric potential is equal to $V_b$ in the background region and is equal to $V_1$ or $V_2$ in each scattering center (disk) sublattice. A periodic spatially varying magnetic field $\mathbf{B}(\mathbf{r}) = \nabla \times \mathbf{A}$ determined by the vector potential

$$\mathbf{A}(\mathbf{r}) = \frac{3B_0 a^2}{16\pi^2}\left[\mathbf{b}_1 \sin(\mathbf{b}_1 \cdot \mathbf{R}) + \mathbf{b}_2 \sin(\mathbf{b}_2 \cdot \mathbf{R}) + (\mathbf{b}_1 + \mathbf{b}_2)\sin([\mathbf{b}_1 + \mathbf{b}_2] \cdot \mathbf{R})\right] \times \hat{\mathbf{z}}, \quad (5)$$

is also applied to the system. Here $a$ is the distance between the nearest scattering centers in the hexagonal lattice, $B_0$ is the peak magnetic field in Tesla, $\mathbf{b}_1$ and $\mathbf{b}_2$ are the reciprocal lattice primitive vectors and $\mathbf{R} = \mathbf{r} - \mathbf{r}_c$ where $\mathbf{r}_c = 1/3(\mathbf{a}_1 + \mathbf{a}_2)$ determines the coordinates of the honeycomb cell's center [Fig. 1a]. The two scatterers are centered at $\mathbf{r}_{0,1} = -a\hat{\mathbf{x}}$ and $\mathbf{r}_{0,2} = \mathbf{0}$, respectively, and have radii $R_1$ and $R_2$.

The stationary states of the (spinless) electronic system are the solutions of the time-independent scalar Schrödinger equation:

$$\left[ \frac{-\hbar^2}{2m_b}\left(\nabla + i\frac{e}{\hbar}\mathbf{A}(\mathbf{r})\right)^2 + V(\mathbf{r}) - E \right]\Psi = 0, \tag{6}$$

where $\Psi$ is the wavefunction, $m_b$ is the electron effective mass, $e$ is the elementary charge and $\hbar$ is the Planck constant.

### B. Photonic analogue of the Haldane model

It was shown in Refs. [23, 27] that a pseudo-Tellegen coupling is the counterpart for photons of the magnetic field coupling for electrons. Specifically, consider a bianisotropic material described by constitutive relations of the type:

$$\begin{pmatrix} \mathbf{D} \\ \mathbf{B} \end{pmatrix} = \mathbf{M} \cdot \begin{pmatrix} \mathbf{E} \\ \mathbf{H} \end{pmatrix}, \qquad \text{with } \mathbf{M} = \begin{pmatrix} \varepsilon_0 \bar{\varepsilon} & \frac{1}{c}\bar{\xi} \\ \frac{1}{c}\bar{\zeta} & \mu_0 \bar{\mu} \end{pmatrix}, \tag{7}$$

where $\bar{\varepsilon}$, $\bar{\mu}$, $\bar{\xi}$ and $\bar{\zeta}$ are the relative permittivity, permeability and magnetoelectric tensors, respectively. The pseudo-Tellegen response is determined by traceless symmetric magneto-electric tensors [28]. In this article, we assume that the relevant tensors are of the form:

$$\bar{\xi} = \bar{\zeta} = \boldsymbol{\xi} \otimes \hat{\mathbf{z}} + \hat{\mathbf{z}} \otimes \boldsymbol{\xi}, \tag{8}$$

where $\boldsymbol{\xi} = \xi_x \hat{\mathbf{x}} + \xi_y \hat{\mathbf{y}}$ is a generic vector lying in the *xoy* plane. We will refer to $\boldsymbol{\xi}$ as the pseudo-Tellegen vector. Some anti-ferromagnets such as Cr$_2$O$_3$ have a Tellegen-type response, albeit it is typically very weak [29-31]. Furthermore, it has been recently shown that some electronic topological insulators may be characterized by a Tellegen type (axion) response [32-35]. Materials with a Tellegen response are nonreciprocal and enable peculiar effects and exotic physics [36-39]. It is interesting to point out that the Tellegen coupling is real-valued. This means that any homogeneous Tellegen material is certainly topologically trivial. Furthermore, it is worth noting that most of the solutions that yield nontrivial photonic topologies rely on gyrotropic materials with a complex-valued material response, very different from the Tellegen case.

Suppose that the relativity permittivity and permeability tensors are of the form

$$\begin{aligned}\overline{\varepsilon} &= \varepsilon_\parallel \left( \hat{\mathbf{x}} \otimes \hat{\mathbf{x}} + \hat{\mathbf{y}} \otimes \hat{\mathbf{y}} \right) + \varepsilon_{zz} \hat{\mathbf{z}} \otimes \hat{\mathbf{z}} \\ \overline{\mu} &= \mathbf{1}_{3\times 3}\end{aligned}, \quad \text{with} \quad \varepsilon_{zz} = 1 - \frac{\omega_p^2}{\omega^2}. \quad (9)$$

Then, it can be shown that the wave propagation of transverse electric (TE) waves (with $E_z \neq 0$ and $H_z = 0$) in a (possibly inhomogeneous) photonic system described by the constitutive relations (7) is ruled by the following wave equation [23, 27]:

$$\left[ -\left( \nabla - i\frac{\omega}{c} \hat{\mathbf{z}} \times \boldsymbol{\xi}(\mathbf{r}) \right)^2 + \frac{\omega_p^2(\mathbf{r})}{c^2} - \frac{\omega^2}{c^2} \right] E_z = 0. \quad (10)$$

It is implicit that the system is invariant to translations along the *z*-direction and that $\partial/\partial z = 0$. As discussed in Ref. [23], the solutions of Eq. (10) can be transformed into the solutions of Eq. (6) using the mapping:

$$\begin{aligned}\frac{2m_b E}{\hbar^2} &\to \frac{\omega^2}{c^2} \\ \frac{2m_b V(\mathbf{r})}{\hbar^2} &\to \frac{\omega_p^2(\mathbf{r})}{c^2} \\ \frac{e}{\hbar} \mathbf{A}(\mathbf{r}) &\to \frac{\omega}{c} \boldsymbol{\xi}(\mathbf{r}) \times \hat{\mathbf{z}}\end{aligned} \quad . \quad (11)$$

This property implies that the Haldane model can be realized in a pseudo-Tellegen photonic crystal. In particular, from Eq. (5) the pseudo-Tellegen coupling must be of the form:

$$\xi(\mathbf{r}) = \xi_0 \frac{\sqrt{3}a}{4\pi}\left[\mathbf{b}_1 \sin(\mathbf{b}_1 \cdot \mathbf{R}) + \mathbf{b}_2 \sin(\mathbf{b}_2 \cdot \mathbf{R}) + (\mathbf{b}_1 + \mathbf{b}_2)\sin([\mathbf{b}_1 + \mathbf{b}_2] \cdot \mathbf{R})\right], \quad (12)$$

where $\xi_0$ is the (dimensionless) peak amplitude of the pseudo-Tellegen vector [23]. Note that the pseudo-Tellegen coupling varies continuously in space, which implies that the dielectric axes that diagonalize the magneto-electric tensors must be space dependent. Furthermore, the spatially varying electric potential can be mimicked by tailoring the electric response ($\varepsilon_{zz}$), i.e., by tailoring the plasma frequency $\omega_p$. As illustrated in Fig. 1b, the plasma frequency associated with the scattering centers ($\omega_{p,i}$, $i=1,2$) is different from the plasma frequency of the background region ($\omega_{p,b} > 0$). The objective of this work is to characterize the topological phases of the photonic Haldane graphene [Eq. (10)] without using a tight-binding approximation.

### III. Topological classification with the Green's function

In a recent series of works [20, 24, 25], we introduced a general Green's function formalism to calculate the gap Chern numbers of non-Hermitian and possibly dispersive photonic crystals. In its most general form, the spectrum of the system under study is determined by a generic differential operator $\hat{L}_\mathbf{k}$ (which is not required to be Hermitian) and by a multiplication (matrix) operator $\mathbf{M}_g$, which determine a generalized eigenvalue problem $\hat{L}_\mathbf{k} \cdot \mathbf{Q}_{n\mathbf{k}} = \mathcal{E}_{n\mathbf{k}} \mathbf{M}_g \cdot \mathbf{Q}_{n\mathbf{k}}$. Here, $\mathbf{Q}_{n\mathbf{k}}$ are the Bloch modes envelopes with the real wave vector $\mathbf{k} = k_x \hat{\mathbf{x}} + k_y \hat{\mathbf{y}}$ and $\mathcal{E}_{n\mathbf{k}}$ are the (generalized) eigenvalues.

Suppose that $\mathcal{E}_{gap}$ is some "energy" lying in the spectral band gap of interest. The gap Chern number $\mathcal{C}_{gap}$ can be found from "first principles" by integrating the system Green's function in the complex frequency plane. The Green's function is defined by $\mathcal{G}_\mathbf{k}(\mathcal{E}) = i(\hat{L}_\mathbf{k} - \mathbf{M}_g \mathcal{E})^{-1}$. The gap Chern number can be conveniently expressed as [13, 24 25, 26]:

$$\mathcal{C}_{gap} = \frac{i}{(2\pi)^2} \iint\limits_{B.Z.} d^2\mathbf{k} \int\limits_{\mathcal{E}_{gap}-i\infty}^{\mathcal{E}_{gap}+i\infty} d\mathcal{E}\, \text{Tr}\{\partial_1 \hat{L}_\mathbf{k} \cdot \mathcal{G}_\mathbf{k} \cdot \partial_2 \hat{L}_\mathbf{k} \cdot \mathcal{G}_\mathbf{k} \cdot \mathbf{M}_g \cdot \mathcal{G}_\mathbf{k}\}, \qquad (13)$$

where $\text{Tr}\{...\}$ stands for the trace operator and $\partial_j = \partial/\partial k_j$ ($j$=1,2) with $k_1 = k_x$ and $k_2 = k_y$. The integral in $\mathcal{E}$ is along a vertical line in the complex plane ($\text{Re}\{\mathcal{E}\} = \mathcal{E}_{gap}$) parallel to the imaginary frequency (energy) axis and completely contained in the bandgap. The integral in $\mathbf{k}$ is over the first Brillouin zone. As further discussed later, the operators $\hat{L}_\mathbf{k}, \mathbf{M}_g$ may be approximated by finite rank matrices that represent the operators in a truncated plane wave basis. The integrations can be performed using standard numerical quadrature rules [25]. For further details on the numerical aspects of the method the reader is referred to Ref. [25]. Different first principles methods for the Chern number calculation are reported in Ref. [40, 41].

A.     Gap Chern number for the electronic Haldane model

It is instructive and pedagogical to apply the Green's function method to the electronic Haldane model (tight-binding approximation). In this case, one takes $\mathbf{M}_g = \mathbf{1}_{2\times 2}$ as the identity matrix and $\hat{L}_\mathbf{k} = \mathbf{H}(\mathbf{k})$ with $\mathbf{H}(\mathbf{k})$ defined as in Eq. (1). Hence, the gap Chern number is simply given by:

$$\mathcal{C}_{gap} = \frac{1}{2\pi} \iint\limits_{B.Z.} d^2\mathbf{k}\, \mathcal{F}_\mathbf{k}. \qquad (14a)$$

$$\mathcal{F}_\mathbf{k} = \frac{1}{2\pi} \int\limits_{\mathcal{E}_{gap}-i\infty}^{\mathcal{E}_{gap}+i\infty} d\mathcal{E}\, \text{Tr}\left\{\frac{\partial \mathbf{H}}{\partial k_1} \cdot (\mathbf{H} - \mathbf{1}_{2\times 2}\mathcal{E})^{-1} \cdot \frac{\partial \mathbf{H}}{\partial k_2} \cdot (\mathbf{H} - \mathbf{1}_{2\times 2}\mathcal{E})^{-2}\right\}. \qquad (14b)$$

The function $\mathcal{F}_\mathbf{k}$ gives the total Berry curvature of the bands below the gap expressed in terms of the Green's function. For the electronic Haldane model the gap energy can be taken equal to $\mathcal{E}_{gap}=0$. Note that in this example the trace operator acts on a $2\times 2$ matrix.

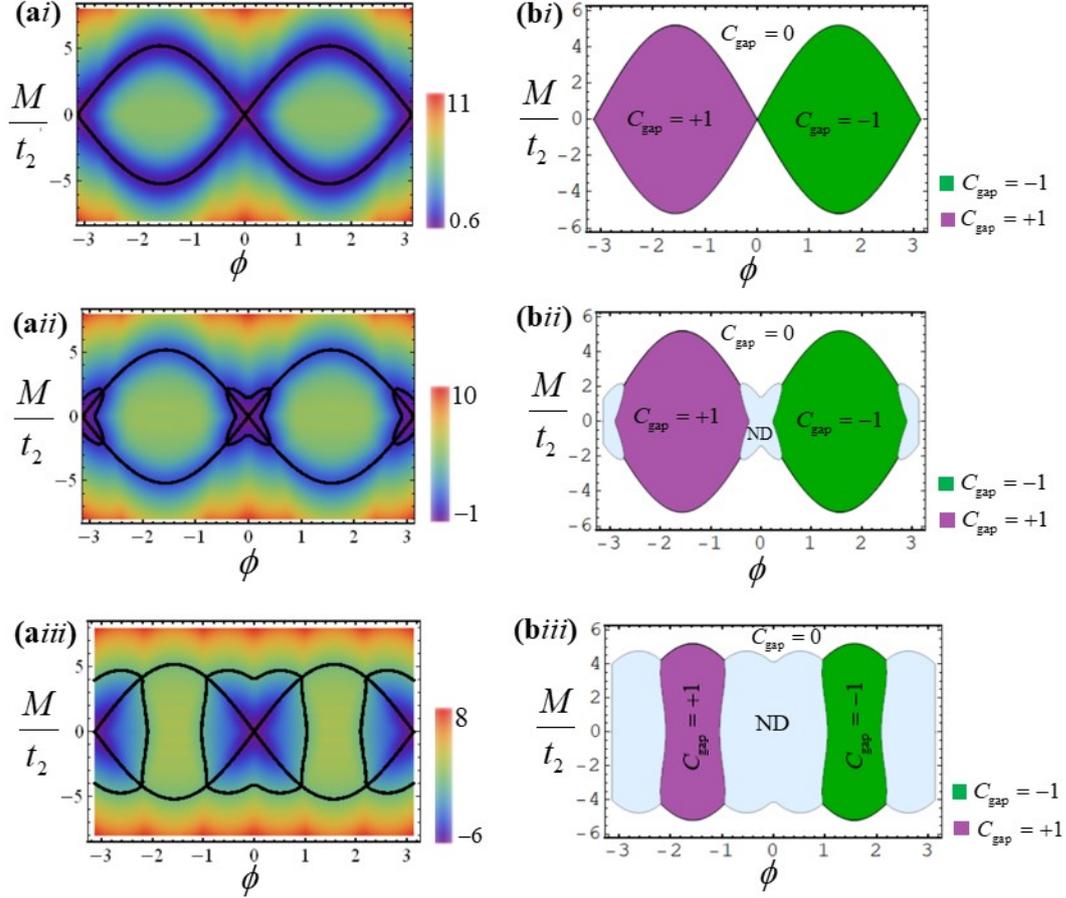

**Fig. 2 (a)** Density plot of the band-gap energy for different combinations of the phase and mass parameters $(\phi, M)$, for $t_1=-3t_2$, (ii) $t_1=-2.5t_2$, and (iii) $t_1=-0.5t_2$. The thick black lines correspond to the pairs $(\phi, M)$ where the band-gap closes. Negative values of the band-gap energy represent configurations for which there is no band-gap. **(b)** Numerically calculated topological phase diagram for the same configurations as in (a). The gap Chern number $\mathcal{C}_{gap}$ is determined through Eq. (13). The different topological phases are shaded with different colors. The ND (non-defined) region corresponds to the pairs $(\phi, M)$ for which there is no gap, and thereby the topological number is not defined.

The gap Chern number was numerically determined for different combinations of $(\phi, M)$. In this manner, we recovered the well-known phase diagram represented in Fig. 2bi).

As it is well established, the system is characterized by a non-trivial topological phase ($C_{gap} = \pm 1$) when the effects of the broken time-reversal symmetry dominate. Otherwise, when the effects of the broken inversion symmetry dominate (large $|M|$ parameter) the system is topologically trivial. Note that when $\phi \neq 0, \pi$ the system has a broken time-reversal symmetry, while when $M \neq 0$ the system has a broken inversion symmetry.

In Fig. 2a, we show a density plot of the bandgap width $\mathcal{E}_{gap}$ in between the two energy bands of the Haldane model for different values of $(\phi, M)$. The band-gap width is defined as $\mathcal{E}_{gap} = \min_{\mathbf{k} \in BZ} \mathcal{E}_{\mathbf{k}}^{+} - \max_{\mathbf{k} \in BZ} \mathcal{E}_{\mathbf{k}}^{-}$, where $\mathcal{E}_{\mathbf{k}}^{-} < \mathcal{E}_{\mathbf{k}}^{+}$ give the dispersion of the two energy bands. The gap energy $\mathcal{E}_{gap}$ vanishes for the pairs of $(\phi, M)$ where the bandgap closes at the Dirac points corresponding to the black thick lines in Fig. 2. Positive values of $\mathcal{E}_{gap}$ correspond to configurations with a full band-gap, whereas negative values of $\mathcal{E}_{gap}$ correspond to configurations with no band-gap. When $\mathcal{E}_{gap} < 0$, the maximum of the low-energy band is larger than the minimum of the high-frequency band. Note that this situation can occur even if the two bands do not intersect and $\mathcal{E}_{\mathbf{k}}^{-} < \mathcal{E}_{\mathbf{k}}^{+}$ in the entire Brillouin zone. As seen in Fig. 2a, it is possible to have configurations with $\mathcal{E}_{gap} < 0$ when $|t_1|/t_2$ is sufficiently small (panels ii) and iii)). To our best knowledge, the case $\mathcal{E}_{gap} < 0$ was not previously discussed in the literature. Figure 2b shows the numerically calculated topological phase diagram, with the gap Chern number obtained by numerical integration [Eq.(13)]. Evidently, the Chern number is not defined in the regions with $\mathcal{E}_{gap} < 0$ (ND regions). Due to this reason, the diagrams of Figs. 2bii) and 2biii) are different from the standard diagram of Fig. 2bi).

Figure 3 shows the density plots of the Berry curvature numerically calculated with the Green's function theory [Eq. (14b)] for different combinations of the $\phi$ and $M$ parameters.

The Brillouin zone is parameterized as $\mathbf{k} = \beta_1 \mathbf{b}_1 + \beta_2 \mathbf{b}_2$ with $-1/2 < \beta_i < 1/2$. As seen, for the topologically trivial phase with $M/t_2 = 5$, $\phi = 0$, [Fig. 3ai)] the Berry curvature has both positive and negative values, so that its integral vanishes [Fig. 2bi)]. In contrast, for the topologically nontrivial case with $M = 0, \phi \neq 0$, the Berry curvature is either mostly negative with $\phi = 0.2$ [Fig. 3aii)] or positive with $\phi = -0.2$ [Fig. 3aiii)]. The Berry curvature is peaked near the Dirac points. The Berry curvature can change appreciably within the same topological phase but its integral over the Brillouin zone (the gap Chern number) is invariant.

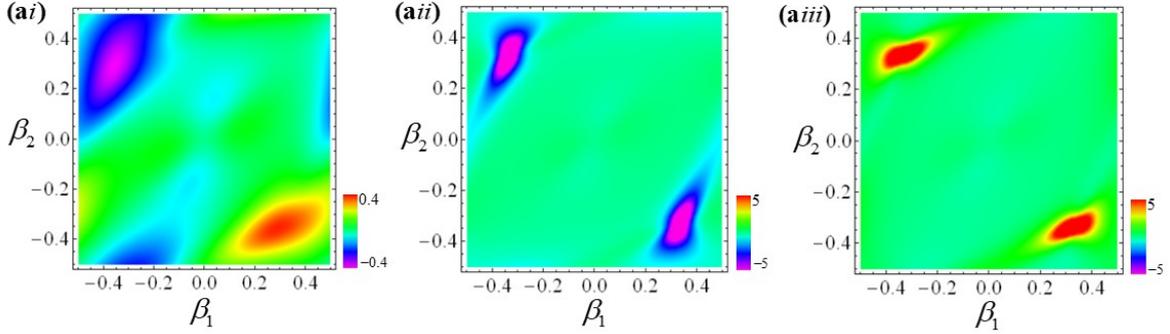

**Fig. 3** Density plot of the Berry curvature of the band below the gap, $\mathcal{F}_{\mathbf{k}}$ [Eq. (14b)] in the first Brillouin zone for $t_1 = -3t_2$. The parameters of the Haldane model are **(ai)** $\phi = 0$ and $M/t_2 = 5$. **(aii)** $\phi = 0.2$ and $M = 0$. **(aii)** $\phi = -0.2$ and $M = 0$.

## B. Topological phases of the photonic Haldane model: Theory

Next, we tackle the more difficult problem of finding the gap Chern number of the photonic Haldane model, without relying on the tight-binding approximation.

To begin with, we rewrite secular equation (10) as

$$\hat{L}(-i\nabla) \cdot E_z = \mathcal{E} E_z \qquad (15)$$

with $\mathcal{E} = (\omega/c)^2$ and,

$$\hat{L}(-i\nabla) = -\left(\nabla - i\frac{\omega}{c}\hat{\mathbf{z}}\times\boldsymbol{\xi}(\mathbf{r})\right)^2 + \frac{\omega_p^2(\mathbf{r})}{c^2}$$

$$\approx -\left(\nabla - i\frac{\omega_0}{c}\hat{\mathbf{z}}\times\boldsymbol{\xi}(\mathbf{r})\right)^2 + \frac{\omega_p^2(\mathbf{r})}{c^2}, \qquad (16)$$

$$= -\nabla^2 + i\frac{\omega_0}{c}\left[\nabla\cdot\left(\hat{\mathbf{z}}\times\boldsymbol{\xi}(\mathbf{r})\right) + \hat{\mathbf{z}}\times\boldsymbol{\xi}(\mathbf{r})\cdot\nabla\right] + \mathcal{V}(\mathbf{r})$$

where $\mathcal{V}(\mathbf{r}) = \frac{\omega_p^2(\mathbf{r})}{c^2} + \frac{\omega_0^2}{c^2}|\boldsymbol{\xi}(\mathbf{r})|^2$. In the second identity of the above equation, we replaced $\omega \to \omega_0$, where $\omega_0$ is some constant frequency. This approximation transforms the secular equation into a standard eigenvalue problem, and avoids complications due to material dispersion. Provided that $\omega_0$ is some frequency inside the relevant gap of the unperturbed operator, the replacement $\omega \to \omega_0$ does not change the physics.

The Bloch modes associated with the wave vector $\mathbf{k} = k_x\hat{\mathbf{x}} + k_y\hat{\mathbf{y}}$ are of the form $E_z = e_z(x,y)e^{i\mathbf{k}\cdot\mathbf{r}}$, with the envelope $e_z(x,y)$ being a periodic function that satisfies

$$\hat{L}_\mathbf{k}\cdot e_z = \mathcal{E}e_z, \qquad \text{with} \qquad \hat{L}_\mathbf{k} = \hat{L}(-i\nabla + \mathbf{k}). \qquad (17)$$

Thus, we obtain a standard eigenvalue problem with a trivial $\mathbf{M}_g = \mathbf{1}$ operator.

In order to find the spectrum of $\hat{L}_\mathbf{k}$ and the topological phases, next we obtain a representation of $\hat{L}_\mathbf{k}$ in a plane wave basis [42]. First, we expand the periodic functions $\boldsymbol{\xi}$ and $\mathcal{V}$ into a Fourier series as $\boldsymbol{\xi} = \sum_\mathbf{J}\boldsymbol{\xi}_\mathbf{J}e^{i\mathbf{G}_\mathbf{J}\cdot\mathbf{r}}$ and $\mathcal{V} = \sum_\mathbf{J}\mathcal{V}_\mathbf{J}e^{i\mathbf{G}_\mathbf{J}\cdot\mathbf{r}}$. Here, $\mathbf{G}_\mathbf{J} = j_1\mathbf{b}_1 + j_2\mathbf{b}_2$ and $\mathbf{J} = (j_1, j_2)$ is a pair of integers. From Eq. (12), it is readily seen that:

$$\boldsymbol{\xi}_\mathbf{J} = \xi_0\frac{\sqrt{3}a}{4\pi}\frac{e^{-i\mathbf{G}_\mathbf{J}\cdot\mathbf{r}_c}}{2i}\begin{cases}\pm\mathbf{b}_1, & \mathbf{J} = \pm(1,0) \\ \pm\mathbf{b}_2, & \mathbf{J} = \pm(0,1) \\ \pm(\mathbf{b}_1 + \mathbf{b}_2), & \mathbf{J} = \pm(1,1) \\ 0, & \text{otherwise}\end{cases}. \qquad (18)$$

The Fourier coefficients of the function $g = \omega_p^2(\mathbf{r})/c^2$ are $p_{g,\mathbf{I}} = \frac{1}{A_{cell}} \int_{cell} g(\mathbf{r}) e^{-i\mathbf{G}_\mathbf{I} \cdot \mathbf{r}} d^2\mathbf{r}$. A straightforward analysis shows that:

$$p_{g,\mathbf{I}} = g_b \delta_{\mathbf{I},0} + \sum_{i=1,2} f_{V,i} (g_i - g_b) e^{-i\mathbf{G}_\mathbf{I} \cdot \mathbf{r}_{0,i}} \frac{2 J_1(|\mathbf{G}_\mathbf{I}| R_i)}{|\mathbf{G}_\mathbf{I}| R_i}, \quad (19)$$

where $\delta_{\mathbf{I},0}$ is Kronecker's symbol, $J_1$ is the cylindrical Bessel function of the first kind and first order, $R_i$ is the radius of the scattering centers of the $i$-th array, $\mathbf{r}_{0,i}$ gives the position of the $i$-th scattering center in the unit cell, and $f_{V,i} = \pi R_i^2 / A_{cell}$ with $A_{cell} = |\mathbf{b}_1 \times \mathbf{b}_2|$ the area of the unit cell. We use $g_b = \omega_{p,b}^2 / c^2$ and $g_i = \omega_{p,i}^2 / c^2$ ($i$=1, 2), with the different $\omega_p$ defined as in Fig. 1. Finally, from $\mathcal{V}(\mathbf{r}) = \frac{\omega_p^2(\mathbf{r})}{c^2} + \frac{\omega_0^2}{c^2} |\xi(\mathbf{r})|^2$, one finds that the Fourier coefficients of $\mathcal{V}$ are given by:

$$\mathcal{V}_\mathbf{J} = p_{\omega_p^2/c^2, \mathbf{J}} + \frac{\omega_0^2}{c^2} \sum_{\substack{\mathbf{J}'=\pm(1,0),\\ \pm(0,1), \pm(1,1)}} \xi_{\mathbf{J}-\mathbf{J}'} \cdot \xi_{\mathbf{J}'}. \quad (20)$$

Here, $p_{\omega_p^2/c^2, \mathbf{J}}$ is defined as in Eq. (19) now with $g = \omega_p^2/c^2$. Note that the second term can be nonzero only for $|j_1| \leq 2$ and $|j_2| \leq 2$.

We are now ready to obtain a representation of $\hat{L}_\mathbf{k}$ in a plane wave basis. To this end, the electric field envelope ($e_z = E_z e^{-i\mathbf{k} \cdot \mathbf{r}}$) is expanded into plane waves as $e_z = \sum_\mathbf{J} c_\mathbf{J}^E e^{i\mathbf{G}_\mathbf{J} \cdot \mathbf{r}}$. Then, it is simple to check that Eq. (17) implies that for a generic double index $\mathbf{I} = (i_1, i_2)$ one has:

$$(\mathbf{k} + \mathbf{G}_\mathbf{I}) \cdot (\mathbf{k} + \mathbf{G}_\mathbf{I}) c_\mathbf{I}^E + \sum_\mathbf{J} \mathcal{V}_{\mathbf{I}-\mathbf{J}} c_\mathbf{J}^E$$
$$-\frac{\omega_0}{c} \sum_\mathbf{J} \left[ (\mathbf{k} + \mathbf{G}_\mathbf{I}) \cdot (\hat{\mathbf{z}} \times \xi_{\mathbf{I}-\mathbf{J}}) + (\hat{\mathbf{z}} \times \xi_{\mathbf{I}-\mathbf{J}}) \cdot (\mathbf{k} + \mathbf{G}_\mathbf{J}) \right] c_\mathbf{J}^E = \mathcal{E} c_\mathbf{I}^E \quad (21)$$

From here it is evident that the operator $\hat{L}_\mathbf{k}$ is represented by the matrix $\hat{L}_\mathbf{k} \to [L_{\mathbf{I},\mathbf{J}}]$ with

$$L_{\mathbf{I},\mathbf{J}} = (\mathbf{k}+\mathbf{G}_{\mathbf{I}})\cdot(\mathbf{k}+\mathbf{G}_{\mathbf{I}})\delta_{\mathbf{I},\mathbf{J}} + \mathcal{V}_{\mathbf{I}-\mathbf{J}} - \frac{\omega_0}{c}\left[(\mathbf{k}+\mathbf{G}_{\mathbf{I}})\cdot(\hat{\mathbf{z}}\times\boldsymbol{\xi}_{\mathbf{I}-\mathbf{J}}) + (\hat{\mathbf{z}}\times\boldsymbol{\xi}_{\mathbf{I}-\mathbf{J}})\cdot(\mathbf{k}+\mathbf{G}_{\mathbf{J}})\right]. \quad (22)$$

The operators $\partial_i \hat{L}_{\mathbf{k}}$ are represented by matrices with generic elements

$$\left[\partial_i \hat{L}_{\mathbf{k}}\right]_{\mathbf{I},\mathbf{J}} = 2\hat{\mathbf{u}}_i\cdot(\mathbf{k}+\mathbf{G}_{\mathbf{I}})\delta_{\mathbf{I},\mathbf{J}} - \frac{\omega_0}{c}\left(2\hat{\mathbf{u}}_i\cdot(\hat{\mathbf{z}}\times\boldsymbol{\xi}_{\mathbf{I}-\mathbf{J}})\right), \quad (23)$$

where $\hat{\mathbf{u}}_1 = \hat{\mathbf{x}}$ and $\hat{\mathbf{u}}_2 = \hat{\mathbf{y}}$ are unit vectors along the coordinates axes. In the numerical calculations the plane wave expansion is truncated imposing that $\mathbf{J} = (j_1, j_2)$ with $|j_m| \leq j_{max}$ with $m=1,2$. To conclude, we note that the spectral problem [Eq. (17)] is equivalent to the matrix eigensystem $\left[L_{\mathbf{I},\mathbf{J}}\right]\cdot\left[c_{\mathbf{J}}^E\right] = \mathcal{E}\left[c_{\mathbf{J}}^E\right]$.

## C. Topological phases of the photonic Haldane model: Numerical results

The band structures of two Haldane photonic crystals are plotted in Fig. 4. The scattering centers have a trivial electric response ($\omega_{p,1} = \omega_{p,2} = 0$), whereas the background region is characterized by the plasma frequency $\omega_{p,b}\, a/c = 5.63$. In the numerical simulations, the plane wave expansion was truncated with $j_{max} = 3$.

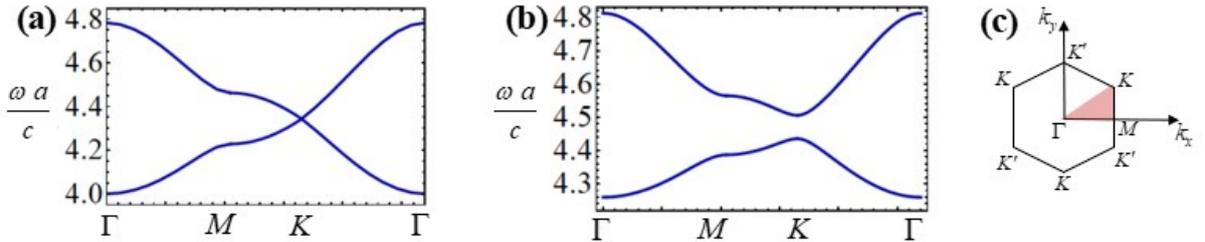

**Fig. 4 (a)-(b)** Band structure of the Haldane photonic crystal. The plasma frequencies of the scatterers are $\omega_{p,1} = \omega_{p,2} = 0$; the plasma frequency of the background material is $\omega_{p,b}\, a/c = 5.63$. The radius of the two scatterers is $R_1 = R_2 = 0.35a$. **(a)** Reciprocal system, with $\xi_0 = 0$. **(b)** Nonreciprocal system with $\xi_0 = 0.677$ and $\omega_0\, a/c = 4.38$. **(c)** Sketch of the first Brillouin zone of the hexagonal lattice.

The reciprocal case characterized by $\xi_0 = 0$ is shown in Fig. 4a, and corresponds to a photonic analogue of graphene [23]. Due to the plasmonic response of the host medium there is a bandgap for low frequencies ($\omega a/c < 4.25$). The nonreciprocal crystal in Fig. 4b is characterized by a nontrivial spatially dependent pseudo-Tellegen response with a peak amplitude $\xi_0 = 0.677$.

As seen in Fig. 4a, for the reciprocal case, the first two bands touch at the Dirac points (*K*) due the symmetry of the hexagonal lattice ($\omega_{p,1} = \omega_{p,2}$). On the other hand, when the pseudo-magnetic field associated with the Tellegen coupling is nontrivial ($\xi_0 \neq 0$) the degeneracy around the Dirac points is lifted, leading to the separation of the bands and to a complete photonic band-gap. Thus, analogous to the electronic Haldane model [2], the time-reversal symmetry breaking opens a gap between the first and the second bands of the system (Fig. 4).

The phase diagram of the Tellegen photonic crystals is plotted in Fig. 5. To model the structural asymmetry between the two sub-lattices of the hexagonal array, we introduce a spatial-asymmetry parameter $\delta$ such that for $\delta > 0$ the plasma frequencies of the scattering centers satisfy $\omega_{p,1} = \delta \times \omega_{p,b}$ and $\omega_{p,2} = 0$, whereas for $\delta < 0$ they satisfy $\omega_{p,1} = 0$ and $\omega_{p,2} = |\delta| \times \omega_{p,b}$. For $\delta \neq 0$ the inversion symmetry of the system is broken. In Fig. 5a we show a density plot of the band-gap "energy" $\mathcal{E}_{gap}$ for different pairs of $(\xi_0, \delta)$. As seen, for the photonic crystal implementation of the Haldane model $\mathcal{E}_{gap} \geq 0$, and so that the topological characterization is feasible in the entire parameter space. The black thick lines represent the combinations of $(\xi_0, \delta)$ for which the band gap closes.

The topological phase diagram of the Tellegen photonic crystal is shown in Fig. 5b. The gap Chern number is calculated from first principles using the Green's function theory. The different topological phases are shaded with different colors and the corresponding gap Chern numbers are identified inside the curves. As seen, analogous to the tight-binding model, when the time-reversal symmetry breaking dominates (large values of $|\xi_0|$), the photonic crystal is topologically nontrivial and the gap Chern number is $\mathcal{C}_{gap} = \pm 1$. In contrast, when the inversion symmetry breaking dominates (large values of $|\delta|$), the system is topologically trivial.

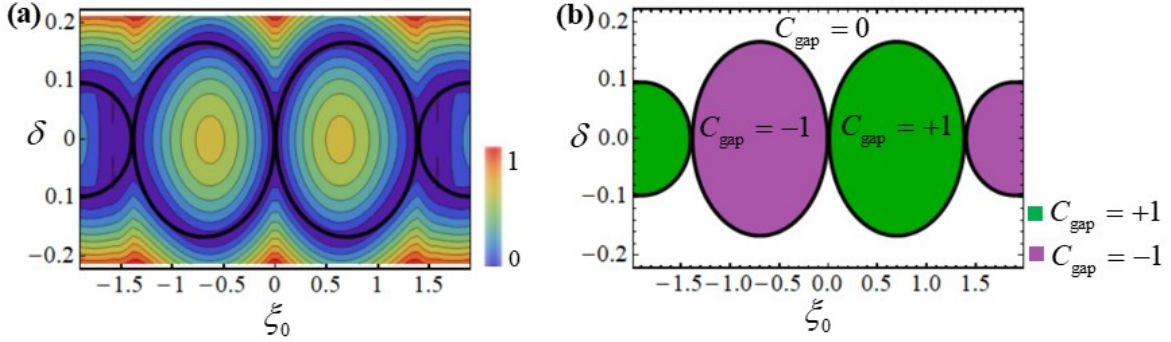

**Fig. 5 (a)** Density plot of the band-gap "energy" $\mathcal{E}_{gap}$ for different combinations of the peak amplitude of the pseudo-Tellegen vector and of spatial-asymmetry parameters $(\xi_0, \delta)$. The thick black lines correspond to the pairs of $(\xi_0, \delta)$ for which the band-gap closes. **(b)** Topological phase diagram of the photonic Haldane photonic crystal. The different topological phases are shaded with different colors. The remaining parameters are as in Fig. 4.

Curiously, the topological charge depends not only on the sign of $\xi_0$, but also on its amplitude. For example, for $0 < \xi_0 < 1.36$ the gap Chern number is +1, but for values of $\xi_0$ slightly larger than 1.36, the sign of the gap Chern number changes due to a phase transition. As a consequence the topological phase diagram is formed by a sequence of bubble-type regions with alternating sign of the gap Chern number. This example illustrates the richness of topological phenomena in photonic crystals, which arises due to periodicity and granular nature of the system, and which can only be captured with a "first principles" approach. It is relevant to note that in a tight-binding approximation it is possible to map the parameters $(\xi_0, \delta)$ into some effective parameters $(\phi, M)$ of the equivalent tight-binding model [Eq.

(1)], see Ref. [22]. Comparing Figs. 2b and 5b, one sees that, for a sufficiently small $|\xi_0|$, the sign of the equivalent $\phi$ is opposite to the sign of $\xi_0$ [22, 23].

## IV. Summary

In this article, we determined the topological phases of a photonic analogue of the Haldane model using a first principles Green's function approach. The proposed system consists of a hexagonal array of dielectric cylinders embedded in a metallic host, with a spatially varying pseudo-Tellegen coupling playing the role a pseudo-magnetic field. The Tellegen nonreciprocal response is at origin of the nontrivial topology of the photonic crystal. Interestingly, the results of the first principles calculations show that even though a bulk pseudo-Tellegen medium has a trivial topology, a nonuniform pseudo-Tellegen structure (photonic crystal) can have topological bandgaps. Furthermore, it was found that due to the complex wave interactions arising from the scattering by the potential-well centers, the phase diagram of the photonic crystal can have nontrivial features and a bubble-type structure different from what is predicted by Haldane's tight-binding model. We expect that the Green's function method can find many other applications in the characterization of the topology of emerging photonic systems.

**Acknowledgements**

This work was supported by the IET under the A F Harvey Engineering Research Prize, the Simons Foundation under the award 733700 (Simons Collaboration in Mathematics and Physics, "Harnessing Universal Symmetry Concepts for Extreme Wave Phenomena"), and by Fundação para Ciência e a Tecnologia (FCT) under project UIDB/50008/2020.